\documentclass[twocolumn]{aastex7}
\usepackage{color}
\usepackage[titletoc]{appendix}
\usepackage{amsmath}
\usepackage{amssymb}
\usepackage{mathtools}
\usepackage{upgreek}
\usepackage{CJK}
\usepackage{comment}
\usepackage{enumitem}
\usepackage{gensymb}
\usepackage{natbib}
\usepackage{graphicx}
\usepackage{bm}
\usepackage{totcount}
\usepackage{multirow}
\usepackage{hyperref}
\usepackage{cleveref}
\usepackage{tabularx}
\usepackage{makecell}
\usepackage[T1]{fontenc}
\usepackage{booktabs}
\usepackage[separate-uncertainty=true]{siunitx}
\usepackage{subcaption}
\usepackage{array}

\newtotcounter{citnum} 
\def\oldbibitem{} \let\oldbibitem=\bibitem
\def\bibitem{\stepcounter{citnum}\oldbibitem}

\crefname{subsection}{subsection}{subsections}

\newcommand{\dd}[2]{\frac{d #1}{d #2}}

\newcommand{\fracbrac}[2]{\left(\frac{#1}{#2}\right)}

\newcommand{\ii}{\mathrm{i}}
\shortauthors{}
\shorttitle{}

\newcommand{\eccRMSE}{0.013}
\newcommand{\incRMSE}{0.0056}

\begin{document} 
\begin{CJK*}{UTF8}{gbsn}
\title{A machine learning model to predict proper orbital elements for main-belt asteroid family classification}

\author[0009-0006-7701-9174]{Yang Shen (沈洋)}
\altaffiliation{Both these authors contributed equally to this manuscript and reserve the right to list themselves as first author on their respective CVs.}
\affiliation{Department of Physics, Harvey Mudd College, Claremont, CA 91711, USA}
\email{lshen@hmc.edu}

\author[0009-0004-8126-3642]{Kavi Dey}
\altaffiliation{Both these authors contributed equally to this manuscript and reserve the right to list themselves as first author on their respective CVs.}
\affiliation{Department of Physics, Harvey Mudd College, Claremont, CA 91711, USA}
\email{kdey@hmc.edu}

\author[0000-0002-1032-0783]{Sam Hadden}
\affiliation{Canadian Institute for Theoretical Astrophysics, 60 St George Str, Toronto, ON M5S 3H8, Canada}
\email{hadden@cita.utoronto.ca}

\author[0000-0001-8736-236X]{Kathryn Volk}
\affiliation{Planetary Science Institute, 1700 E Fort Lowell Road STE 106, Tucson, AZ 85719, USA}
\email{kvolk@psi.edu}

\author[0000-0002-9908-8705]{Daniel Tamayo}
\altaffiliation{Corresponding author: \href{mailto:dtamayo@hmc.edu}{dtamayo@hmc.edu}}
\affiliation{Department of Physics, Harvey Mudd College, Claremont, CA 91711, USA}
\email{dtamayo@hmc.edu}

\begin{abstract}
While fragments from asteroid collisions are born on nearly identical orbits, their shared eccentricity and inclination oscillations get out of phase with one another over time. 
This scrambling makes it hard to identify which fragments belong to the same collisional family when making observations at the present day. 
A powerful tool for overcoming this problem is the calculation of proper orbital elements, which are approximately conserved quantities that would be shared by collisional fragments and provide a transformed space in which to identify clusters of family members.
Traditionally, estimating proper elements for main-belt asteroids requires performing Fourier analysis of multi-Myr numerical integrations, requiring approximately 30 minutes per asteroid with current hardware. Given the order-of-magnitude increase in main-belt asteroids expected by LSST, this approach would require of order 2.5 million CPU hours to assign new detections to known families. To mitigate this computational bottleneck, we present a machine learning model to predict main-belt asteroids' proper elements. Our \texttt{XGBoost} model is trained on the catalogue of proper elements for 1.25 million main-belt asteroids derived through traditional methods by \cite{2024ApJS..274...25N}. The resulting predictions achieve Root Mean Square (RMS) error of $\eccRMSE$ for proper eccentricity predictions and $\incRMSE$ for sin(inclination). We obtain an over $150,000$-fold speedup in estimating proper elements relative to traditional $N$-body methods, which could prove a useful tool for asteroid family identification with the LSST survey. We find that using our predicted proper elements, 95\% of a sample of held out asteroids are correctly assigned to their known family (or to no family for asteroids with no known family).

\end{abstract}

\section{Introduction}
\label{sec: Introduction}
Impacts in the asteroid belt disrupt parent bodies to create families of collisional fragments.
Identifying and characterizing such asteroid families at the present day constrains the belt's collisional history, asteroids' mechanical and chemical properties, and elucidates dynamical pathways to near-Earth space \citep[for recent reviews, see][]{Nesvorny15, Masiero15, Michel15, Novakovic22}.

\begin{figure*}
    \centering
    \includegraphics[trim={0 0.25cm 0 0},clip]{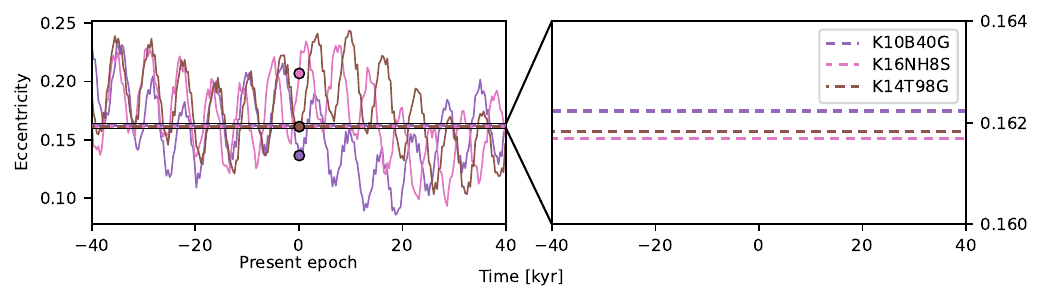}
    \caption{\label{fig:proper_elems_expln} \textit{Left}: Eccentricity evolutions of three asteroids from the Vassar family integrated forward and backward for $40,000$ years from Nesvorn\'y's epoch. The overlapping straight lines indicate the proper eccentricities of these three asteroids. \textit{Right}: The same proper eccentricities, plotted with a rescaled vertical axis to highlight their small differences.}
\end{figure*}

Such identification is complicated by the fact that, while the fragments approximately inherit their parent body's eccentricity and inclination evolution, slight differences between their semimajor axes cause their evolution to get out of phase with one another over time.
In this idealized picture, if one observed several fragments (members of the same asteroid family) at a common epoch at the present day, they would each be at a different phase along their common evolution.
Of course the phase is not known a priori, so all one would see is asteroids with different eccentricities as shown in left panel of Figure\:\ref{fig:proper_elems_expln}, mixed in with many non-family members.
This makes it difficult to identify asteroid families directly.

Instead, analytic and numerical techniques have been developed to obtain so-called ``proper elements''.
By removing oscillatory components of the orbital evolution forced by the giant planets, one can extract a ``proper eccentricity'' that is (approximately) equal for all the family members (left panel of Figure \ref{fig:proper_elems_expln}).
Transforming main belt asteroids' osculating orbital elements to proper elements reveals significant clustering.
Proper elements have thus been used to identify separate asteroid families, historically with hierarchical clustering methods \citep{1990AJ....100.2030Z, 1994AJ....107..772Z, 2014Icar..239...46M, Nesvorny15, 2022CeMDA.134...34N, 2024ApJS..274...25N} and more recently using machine-learning techniques \citep{Carruba2019-iz, carruba2020machine, Lourenco2022-qo, shams2025asteroid}.
This paper is instead focused on faster numerical techniques for estimating the proper elements of main belt asteroids to enable such family identification with current observational surveys.

\cite{2024ApJS..274...25N} numerically determine the proper elements for a sample of $\sim 1.25$ million main-belt asteroids, identifying and removing forced oscillations through frequency-modified Fourier transforms \citep{laskar1993frequency,vsidlichovsky1996frequency} of N-body integrations over the past 10 Myr. This method's numerical evaluation of each asteroid's set of proper elements requires approximately 30 minutes of CPU time (comparable to the 20 minutes per asteroid estimated by \citealt{spencer_volk_ragozzine_malhotra_2026} for similar proper element calculations).
This presents a significant computational bottleneck for such analyses on the Rubin Observatory's expected yield of $\sim 5$ million new main-belt asteroids \citep{jones2020scientific,Kurlander:2025}.

In this paper we train a machine learning model to predict main-belt asteroids' proper elements, using the dataset of $\approx 1.25$ million numerically computed values from \cite{2024ApJS..274...25N} as training examples. 
In Section\:\ref{sec: Machine Learning Model} we describe the setup and training process of our machine learning model.
In Section\:\ref{subsec: Model Performance} we evaluate the model's performance predicting proper elements. In Section\:\ref{subsec: Family Candidate Identification} we evaluate the model's performance on new asteroid family candidates. We release our code (\href{https://github.com/YangShen31/proper_elements_prediction}{github.com/YangShen31/proper\_elements\_prediction}) and trained machine learning model (\href{https://doi.org/10.5281/zenodo.20693707}{doi.org/10.5281/zenodo.20693707}) publicly.

\section{Machine Learning Model}
\label{sec: Machine Learning Model}

We frame this problem as a supervised regression task, trying to predict the proper eccentricity and inclination for a given asteroid without doing a costly $N$-body integration. We began with a dataset of asteroids with known proper elements. For predicting both proper eccentricity and inclination, we construct a feature vector consisting of computationally inexpensive quantities that characterize that asteroid. Then we train a machine learning regression algorithm to predict each asteroid's proper elements from the corresponding feature vector (specifically we predict the difference between the osculating and proper elements). This is a supervised technique as the algorithm learns from real asteroid's feature vector -- proper element pairs during the training process. 
In Section \ref{subsec: Dataset}, we describe our dataset of asteroid  proper elements and the construction of feature vectors and in Section \ref{subsec: Model Training} we describe our regression model.

\subsection{Dataset}
\label{subsec: Dataset}

To construct the dataset used to train the model, we use the proper element catalog calculated by \cite{2024ApJS..274...25N} for $1.25$ million asteroids. The study integrated asteroids under perturbations from the Solar System planets with the \texttt{Swift} orbital integration package \citep{Levison:1994swift} and then removed forced oscillations identified using frequency-modified Fourier Transforms  \citep[FMFT;][]{1993Natur.361..608L, vsidlichovsky1996frequency} to calculate the proper elements (we refer to these proper elements as ``N-Body proper''). The Nesvorn\'y catalog has the majority of its osculating orbits measured at epoch JD2460200.5. Asteroids with osculating elements reported at different epochs constitute only $\sim 1\%$ of the dataset and were excluded for consistency, resulting in a dataset of 1,249,051 asteroids.

We separately train two regression models, one to predict proper eccentricity, the other proper inclination. Each model's feature vector has two parts\footnote{We tested additionally including proper semi-major axes, determined via short $N$-body integrations, as elements of asteroids' feature vectors and found negligible performance improvement. Therefore, we choose to use osculating semi-major axis to avoid additional computation costs.}: 1. a combination of the osculating orbital elements in the invariable frame at epoch JD2460200.5, 2. analytic estimates of the proper elements based on linear Laplace-Lagrange theory.

In order to avoid coordinate singularities for nearly circular / nearly co-planar orbits, the osculating elements are provided as the Cartesian variables
\begin{align}
    \begin{split}
        q&=\sin i \cos \Omega, \\
        p&=\sin i \sin \Omega, \\
        k&=e \cos \varpi, \\
        h&=e \sin \varpi
    \end{split}
\end{align}
where $i$ is the osculating inclination, $\Omega$ the osculating longitude of the ascending node, $e$ the osculating eccentricity, and $\varpi$ the osculating longitude of the pericenter.

The second part of an asteroid's feature vector supplies analytic estimates of proper elements derived from linear Laplace-Lagrange theory \citep[e.g.][]{murray_dermott_2010}. We detail our analytic calculations in Appendix \ref{sec:appendix:linear_theory}. We will refer to these predictions for the proper elements as ``linear predicted'' elements. We add the linear predicted proper eccentricity to the eccentricity model's feature vector (and the corresponding quantity inclination model's feature vector). 
We experimented with providing both linear predictions for the proper eccentricity and inclination models, but the difference in performance is negligible so we opted for the smaller set of inputs to prevent overfitting. 

The \cite{2024ApJS..274...25N} dataset was randomly split into a training set of 80\% of the labeled examples, and a holdout set of 20\% to evaluate performance on unseen data. The training set is further split up into training and validation with cross validation as described in the following section.

\subsection{Model Description}
\label{subsec: Model Training}
We trained and compared several regression algorithms to predict the difference between osculating and N-Body proper elements, including k Nearest Neighbor with scikit-learn \citep{JMLR:v12:pedregosa11a}, pytorch neural networks \citep{3454287.3455008}, \texttt{XGBoost} \citep{10.1145/2939672.2939785}, and \texttt{NGBoost} \citep{2019arXiv191003225D}. We found that \texttt{XGBoost} (gradient boosted decision trees) yielded good RMSE (Root Mean Square Error) deviations from the ground truth, as well as a fast training time for our data. We also tested predicting the value of the proper elements and the difference between osculating and proper elements and found that predicting the difference had better performance. 

\texttt{XGBoost} has a large number of hyperparameters that affect how it trains a model. We find optimal hyperparameters using two grid searches with 5-fold cross validation (CV). For each grid search, we start by generating every combination of hyperparameter values. For each combination, we train 5 models on different subsets of the training data and pick the set of parameters that resulted in the best average performance. The first search finds the optimal values for the model parameters (\texttt{max\_depth}, \texttt{min\_child\_weight}, \texttt{subsample}, \texttt{colsample\_bytree}) and the second search finds optimal training parameters (\texttt{learning\_rate}, \texttt{n\_estimators}). Using two searches decreases the runtime by significantly decreasing the number of hyperparameter combinations to check. During the first search, we use default values for the parameters of the second search (\texttt{learning\_rate} = 0.3, \texttt{n\_estimators} = 500). Most of these hyperparameters act as regularizers that can be tuned to allow for sufficiently flexible models while avoiding overfitting the particular training set. For example, \texttt{n\_estimators} controls the number of trees in the forest, while \texttt{max\_depth} controls the maximum depth of each decision tree (how many `questions' are asked before making a prediction).
The grid search was run separately for the eccentricity and inclination models. Final hyperparameters and model performance can be found in Table \ref{tab:optimal_parameters}. Finding optimal hyperparameters for both models takes 5 hours (wall-clock time) using 40 threads of an AMD Ryzen Threadripper PRO 3975WX. This long time is due to the exhaustive grid search. Once optimal hyperparameters are known training the final models training takes less than a minute.

\begin{table}
    \hspace{-4em}
    \begin{tabular}{l>{\scriptsize}lcc} \toprule
        Parameter & {\small Search Space} & Ecc.& Incl. \\
        \midrule
        \texttt{max\_depth} & $[3,6,9,\dots,21]$ & 9 & 9 \\
        \texttt{min\_child\_weight} & $[1,2,3,4]$ & 1 & 1 \\
        \texttt{subsample} & $[0.8,0.9,1]$ & 1.0 & 1.0 \\
        \texttt{colsample\_bytree} & $[0.8,0.9,1]$ & 1.0 & 0.9 \\
        \texttt{learning\_rate} & $[0.1, 0.01, 0.05]$ & 0.05 & 0.01 \\
        \texttt{n\_estimators} & $[900, 1200, 1500, 2000, 2500]$ & 2500 & 2500 \\
        \midrule
        RMSE & & \eccRMSE & \incRMSE\\
        \bottomrule
    \end{tabular}
    \caption{Parameter search space, and optimal parameters and RMSE found by grid search for eccentricity and inclination models.}
    \label{tab:optimal_parameters}
\end{table}

\begin{figure*}[htbp]
    \centering
    \includegraphics[width=0.9\textwidth]{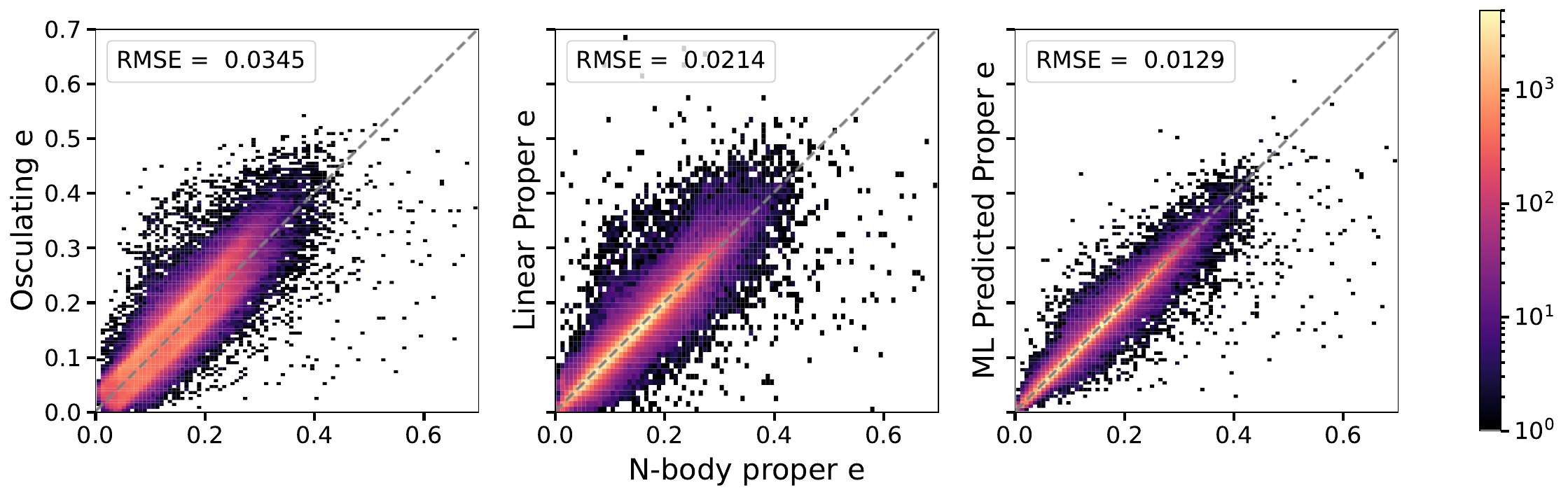}
    \includegraphics[width=0.9\textwidth]{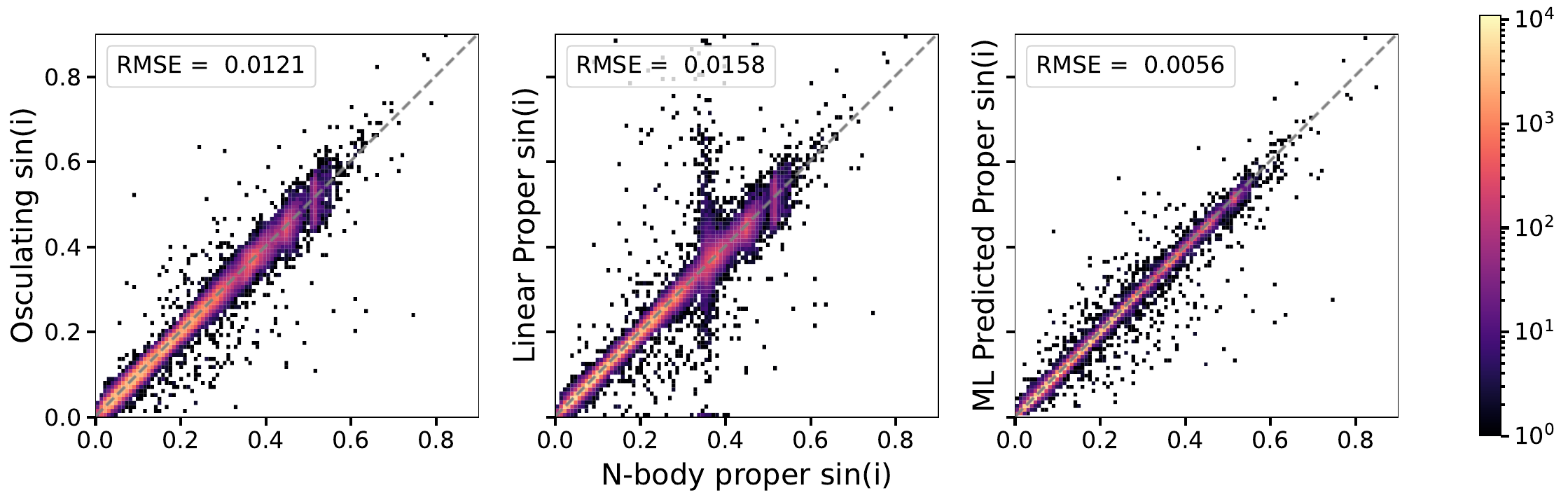}
    \caption{
    Two-dimensional histograms comparing test set asteroids' 
    osculating orbital elements (left), 
    proper elements predicted via linear theory  (middle), 
    and proper elements predicted with our machine learning model (right) 
    against proper elements determined via $N$-body integration.
    The top row shows orbital eccentricities and the bottom row shows inclinations.
    A 1:1 relationship is indicated by a dashed diagonal line in each panel. 
    Each predictor's performance is indicated by an RMSE value with a trend of increasing accuracy evident from left to right (except from osculating to linear predicted $\sin(i)$). 
    }
    \label{fig:model_performance}
\end{figure*}

\section{Results}
\label{sec: Results}

 Evaluating our models requires considering both numerical accuracy in performance and practical applicability. Therefore, we first present the statistical metrics and visualizations of the machine learning model's performance in Section \ref{subsec: Model Performance}. Then, we examine the model's capability to identify asteroid family members in Section \ref{subsec: Family Candidate Identification}. 

\subsection{Statistical Performance Metrics}
\label{subsec: Model Performance}
In Figure\:\ref{fig:model_performance}, we compare the ground-truth, numerically computed proper elements against osculating elements, our linear theory estimates, and the predictions of our machine learning models. As expected, the osculating elements (taken at JD2460200.5) are a poor approximation for the true proper elements. Generally, there is an increasing trend of accuracy from osculating to linear predicted proper, and then to machine learning predicted proper elements. One exception is that the linear predicted $\sin(i)$ has a higher RMSE compared to the osculating one. This is likely to be caused by the spike near the secular resonance at $\sim 0.35$ and outliers. To test this, we clipped the $\sin(i)$ to be below $0.35$ and within $5\sigma$ to the $1:1$ correlation, and we found that the linear $\sin(i)$ predictions improved on the osculating ones by $\sim 30\%$. 
The machine learning models (right-most column) are visibly better, with an $\sim 2-$fold improvement in the RMSE compared to the osculating values.
We note that the colors plot the density of points logarithmically, so the machine learning predictions exhibit a significantly higher density of points close to the 1:1 line. 


\begin{figure*}[htbp]
    \centering
    \includegraphics[width=\textwidth]{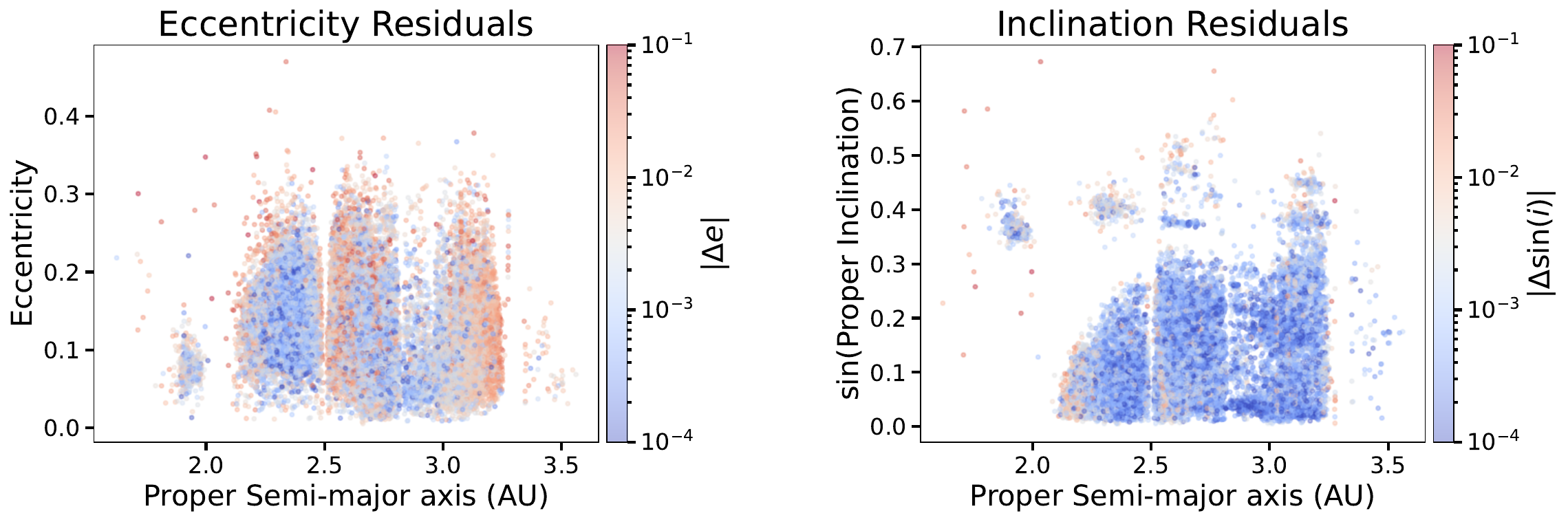}
    \caption{Residuals in proper $e_p$ and $\sin(i_p)$ between the machine learning prediction and the \cite{2024ApJS..274...25N} N-body integrations, following \cite{spencer_volk_ragozzine_malhotra_2026}.}
    \label{fig:dsini_de_plots}
\end{figure*}

Next, we examine the distribution of errors in prediction in proper element space. Figure \ref{fig:dsini_de_plots} plots $\sin(i_p)$ and $e_p$ against $a_p$, with the color scale representing the magnitude of the residuals between the model predicted and N-body calculated proper elements. The residuals are generally small across the main belt, but larger residuals are more frequently found near the edges of the proper space. 
This pattern can be explained by the lower number density in these regions resulting in a smaller training sample, thus causing less accurate prediction. We also find the asteroids with large residuals concentrate near secular and mean-motion resonances (MMRs). The separatrices of the various MMRs and secular resonances are more dynamically unstable and are poorly modeled by the linear theory, resulting in a higher model prediction error. While we do not have a simple explanation for the worse residuals for the proper eccentricity than for the proper inclinations, a possible explanation for the poor performance near the outer edge of the asteroid belt is that it is bounded by the first-order 2:1 eccentricity MMR with Jupiter, which does not exist for the inclinations (only the weaker 4:2 inclination MMR).

\begin{figure*}[htbp]
    \centering
    \includegraphics[width=\textwidth]{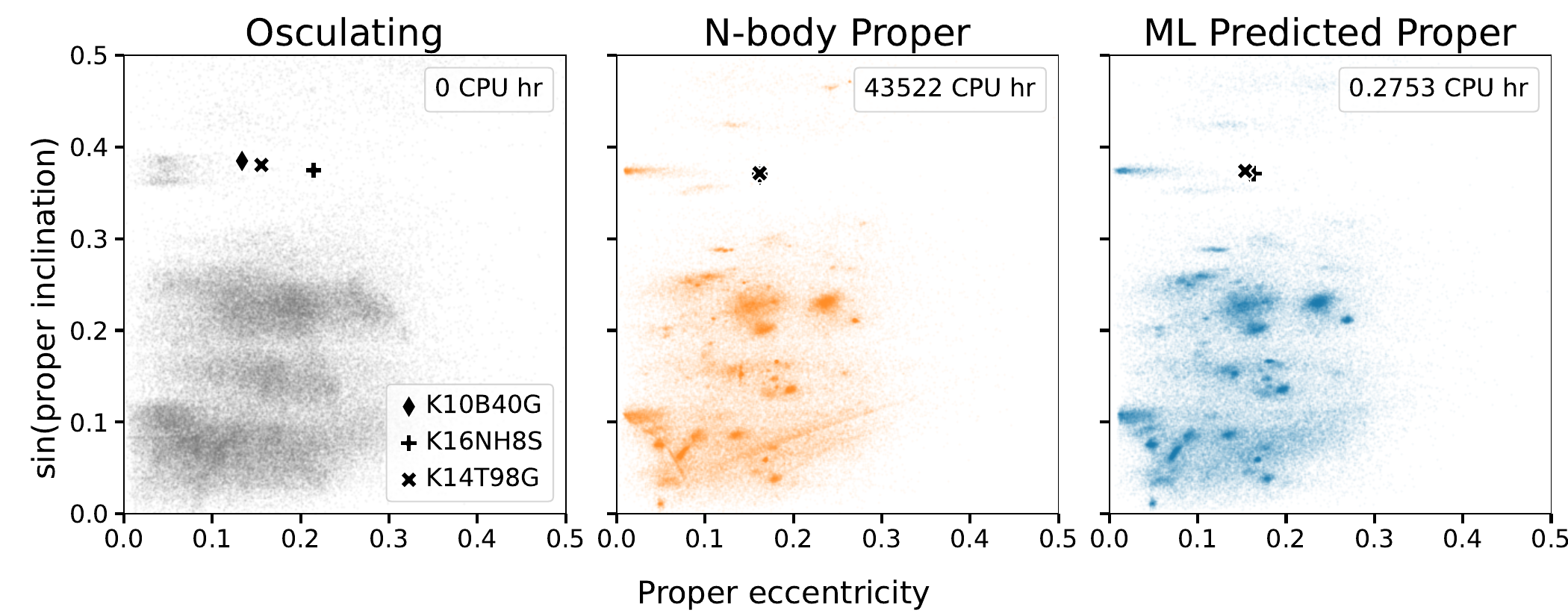}
    \caption{Comparison between osculating, N-body, and machine learning predicted proper elements in the middle asteroid belt. The three asteroids used in Figure \ref{fig:proper_elems_expln} are also plotted in phase space.}
    \label{fig:middle_belt_distribution}
\end{figure*}

One can also try to quantify upper limits on the performance of any model.
The model targets the proper elements of \cite{2024ApJS..274...25N}, each of which has an associated uncertainty; an RMSE of zero is therefore not achievable.
For each asteroid, \cite{2024ApJS..274...25N} calculates proper element uncertainty as the RMSE of the proper elements computed from different intervals across the $10$ Myr integration time span. 
We then perform a Gaussian perturbation test to estimate the scale of deviations in proper elements across intervals. We draw samples from a normal distribution centered on each asteroid's proper element value given by \cite{2024ApJS..274...25N}, with a standard deviation given by the its associated uncertainty. 

Computing the RMSE of these Gaussian perturbations yields \num{6.5d-3} for the eccentricities and \num{3.6d-3} for the inclinations, $50\%$ (eccentricities) and $36\%$ (inclinations) smaller than what our model achieved (see Table~\ref{tab:optimal_parameters}). This provides an estimate of the performance difference between our Machine Learning model and the Nesvorn\'y catalog.

If family members all shared \textit{exactly} the same proper elements, this would be a limitation of the model.
In reality, family members have a small but finite spread in proper elements, so from the standpoint of family classification, we only require that the errors in the model be small compared to the spread in proper elements across typical families.

To explore this, we begin in Figure \ref{fig:middle_belt_distribution} by zooming into a slice of asteroids near the middle of the belt ($2.5 < a< 2.825$ AU), and plot the distribution in $e$ vs. $i$ space.  
We see that in the left panel using osculating values, the large spread among family members washes out any significant structure. 
The middle panel using state-of-the-art N-body methods provides much clearer clustering but takes 43,522 CPU hours (calculated estimating 0.5 hrs/asteroid). 
Finally, with only 0.28 CPU hours of calculation ($\approx 150,000$ times faster, the machine learning predicted proper elements (right panel) provide comparable clustering that visually nearly match those from the N-body proper elements. Total CPU hours is calculated as: time to integrate to JD2460200.5 + linear theory time + model prediction time. 
Linear theory time and model prediction time were measured on average over all 1.25 million asteroids from \cite{2024ApJS..274...25N}. The time to integrate to JD2460200.5 was measured for 258 asteroids as described in Sections~\ref{subsec: Family Identification Pipeline} and Appendix~\ref{subsec:appendix pipeline}. All measurements were taken on an AMD Ryzen Threadripper PRO 3975WX.

The average size of an asteroid family is $\delta_e = 0.012, \delta_{\sin i}=0.0073$. This is smaller than the RMSE between osculating and proper, hence the washed out osculating, but roughly the same size as the RMSE between machine learning predicted and proper elements enabling families to be identified in the machine learning predicted proper phase space.


\begin{figure*}[ht]
    \centering
    \includegraphics[width=\textwidth]{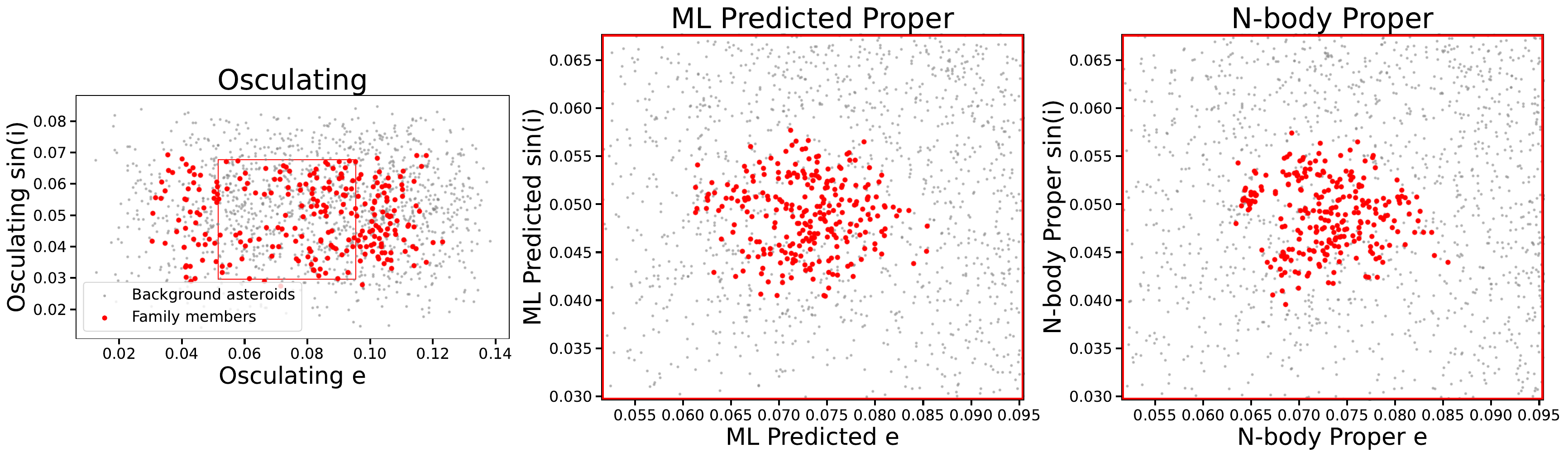}
    \caption{Comparison of Velleda family members distribution in 2-D space of osculating, machine learning predicted proper, and N-body proper orbital elements with osculating semi-major axes $2.32<a<2.47$ AU. The red boxes correspond to the same 2-D projection of the slab in the N-body proper elements space.}
    \label{fig:family_2d}
\end{figure*}
Lastly, we remark that regression models like the one we trained often do not generalize well to out-of-distribution data.
It may therefore be prudent to validate any identification of new families with this model using traditional N-body methods, especially if they lie outside the well sampled regions.
However, we expect the majority of family members discovered by LSST will belong to existing families, where the model's performance should be robust. We also note that LSST will observe smaller and fainter objects, which will be more spread out in phase space due to radiation effects \citep[e.g.,][]{Bottke06}. Our model's performance on currently known asteroids is thus an imperfect proxy. However, we emphasize that even if some applications ultimately require $N$-body-derived proper elements, using rapid tools like this to cut down the candidate list requiring expensive computation by $\sim 2$ orders of magnitude (see discussion around Fig. 7) will provide significant efficiency gains.

\subsection{Family Candidate Identification}
\label{subsec: Family Candidate Identification}
Identifying asteroid families relies on finding clusters of bodies in proper element space (e.g., right panel of Figure\:\ref{fig:middle_belt_distribution}). 
The right panel of Figure\:\ref{fig:family_2d} shows a zoom-in with members of the Velleda family identified by \cite{2024ApJS..274...25N} highlighted in red.
The projection into the 2-D eccentricity-inclination plane renders the semimajor axis of each body invisible, but they have been restricted to a slice of $2.33 < a < 2.46$ AU.

Quantifying how similar two asteroids are requires defining a measure of distance. \cite{2024ApJS..274...25N} adopts the metric $d$ of \cite{1990AJ....100.2030Z},
\begin{equation}
d = {3 \times 10^4\, {\rm m/s} \over \sqrt{a}} \sqrt{ {5 \over 4} 
\left({\delta a/a}\right)^2 + 2 (\delta e)^2 + 2 (\delta \sin i)^2}\ , 
\end{equation}
where $\delta e$, and $\delta \sin(i)$ are differences between orbital elements for a pair of asteroids (osculating or proper depending on which set of elements are being evaluated). $\delta a$ is always the difference between osculating semi-major axis, and $a$ is the average semi-major axis of the two asteroids being compared.

For each family, \cite{2024ApJS..274...25N} identifies a cutoff value $d_\text{cutoff}$ and defines family membership by requiring a candidate asteroid to be within $d_\text{cutoff}$ of at least one other family member. They use a hierarchical clustering algorithm (HCM; \citealt{1990AJ....100.2030Z}) to progressively add candidates as known family members.

Once a family is known, determining whether a new asteroid detection belongs to the family is straightforward---one simply tests whether its proper elements lie within $d_\text{cutoff}$ of a known family member.
We now wish to evaluate our ability to establish family membership using our ML-predicted proper elements. 

As a first step, one can imagine using instead the osculating elements as the outputs of a particularly bad proper element model (left panel of Figure\:\ref{fig:family_2d}, note the significantly zoomed out axes).
If one imagines taking a random red point as a simulated asteroid detection, one would have to adopt a much larger $d_\text{cutoff}$ to correctly associate it with the family. 
But this also means that many background (blue) asteroids would also be falsely classified as family members, leading to a low sample purity or precision (fraction of predicted family members that actually belong). 

This is the tradeoff faced by all models.
The larger $d_\text{cutoff}$ one selects to try to detect a large fraction of all the family members (high recall), the more background asteroids enter, leading to a low precision.
This is improved with a better predictor like our ML model.
In the middle panel we see that the spread of ML-predicted proper elements is slightly larger but comparable to the N-body values in the right panel.
This leads to comparable precision for the same $d_\text{cutoff}$ as in the analysis of \cite{2024ApJS..274...25N}.

\begin{figure}
    \centering
    \includegraphics[width=\columnwidth]{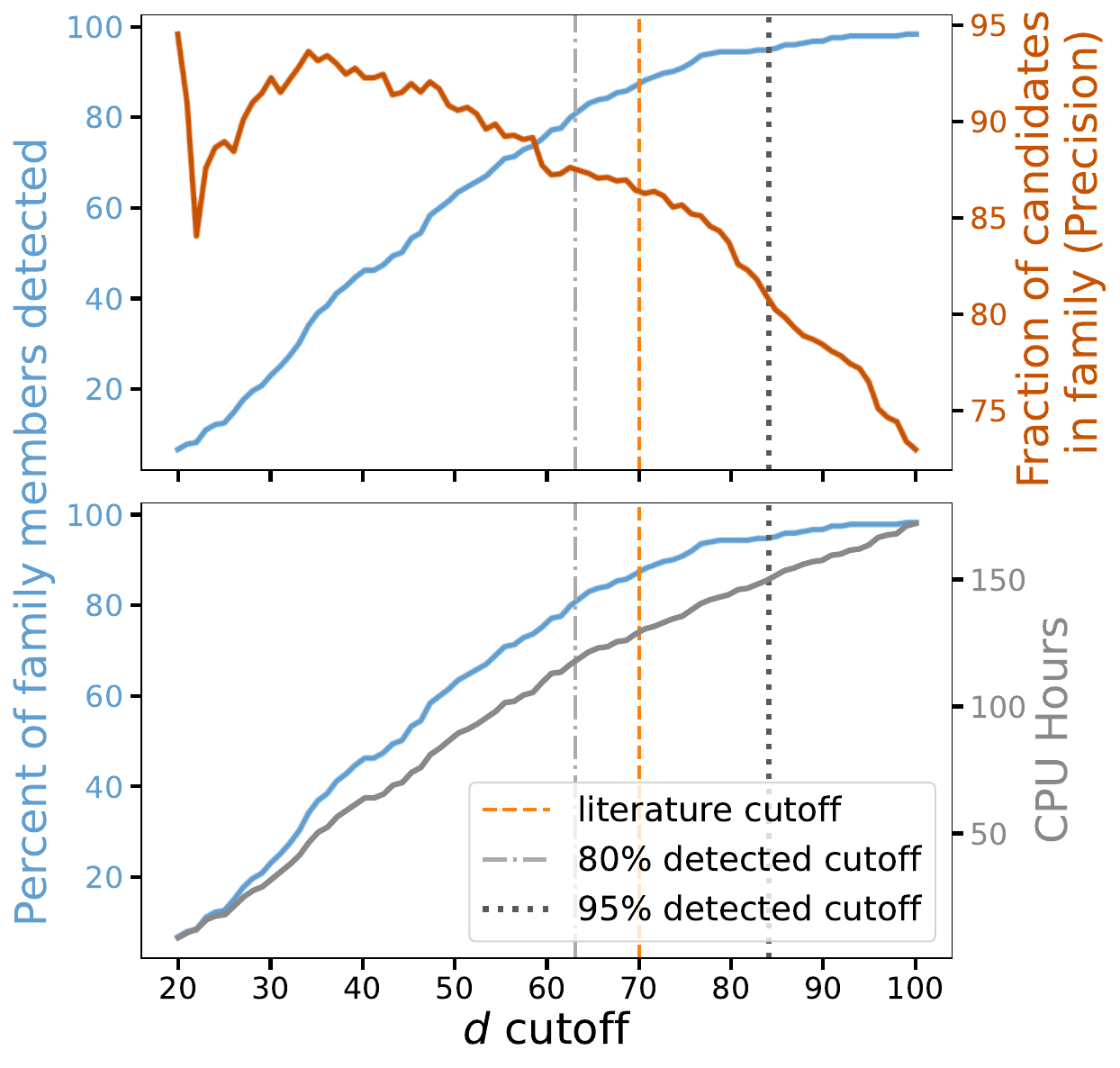}
    \caption{The relation between d values and the percentage of family members detected, the precision of the test sample, and the CPU hours to confirm all detected asteroids for the Velleda family with N-body integrations. The literature $d$ cutoff is obtained from \cite{2024ApJS..274...25N}.}
    \label{fig:Velleda_family}
\end{figure}



To quantify the trade-offs between precision and recall, we evaluate all the train set asteroids within a 3D slab encompassing the known family members identified by \cite{2024ApJS..274...25N} Based on this evaluation, we determine the $d$ cutoffs corresponding to different recalls and subsequently assess them on the test set asteroids. For each simulated asteroid detection, we then use our ML-predicted elements to evaluate the $d$ distance between the candidate and each of the known family members (using their N-body proper elements already known from \citealt{2024ApJS..274...25N}). 
We only consider asteroids in the train set as candidates, using a ``leave one out'' method where each asteroid is compared against all family members \textit{except} itself (for asteroids not in the family, this has no effect).

One example of a 2-D projection of the slab is the square box shown in Figure \ref{fig:family_2d}'s left panel.
Since we are comparing the machine learning predicted elements of train asteroids to the N-body proper elements for known family members, our slab needs to encompass all family member asteroids in both machine learning predicted and N-body proper space. For each dimension (semi-major axis, eccentricity, and inclination) we define the slab boundaries using the minimum and maximum values of the combined model-predicted and N-body proper elements of the known family members, with an additional padding of $\pm 0.01$. 
Since our model does not return predicted proper semi-major axes, we use osculating semi major axes instead for the slab boundaries and later family membership calculations. This ensures we have a representative sample of background asteroids that our model should identify as \emph{not} part of the family, and true family members.

We see in the top panel of Figure\:\ref{fig:Velleda_family} that as one raises $d_\text{cutoff}$ along the $x$-axis, the envelope for family membership in the middle panel of Figure\:\ref{fig:family_2d} expands, and the blue line in Figure\:\ref{fig:Velleda_family} that traces the percentage of red family members recovered (recall) rises.
When $d_\text{cutoff}$ reaches the vertical dashed orange line, showing the canonical value adopted by \cite{2024ApJS..274...25N}, we see that our ML pipeline detects over 87\% of the Velleda family members. The dot-dash and dotted grey lines show the $d_\text{cutoff}$ thresholds which recover 80\% and 95\% of family members respectively. 

The solid blue and brown lines in the top panel also shows the generic tradeoff between precision and recall inherent to any classification model.
As $d_\text{cutoff}$ rises to envelop a larger fraction of family members, more background asteroids also creep in, and the precision (the fraction of predicted family members that actually are in the family) drops in the solid brown line. 

An appropriate $d$ cutoff likely depends in part on the application.
For example, one might want to minimize misclassification of family members to optimize observation time (high precision in solid brown).
Such a strategy will inherently mean a lower recall (solid blue), so some family members will be missed.
But one can always adjust $d$ in either direction to better match one's goals.

\begin{figure}
    \centering
    \includegraphics[width=\columnwidth]{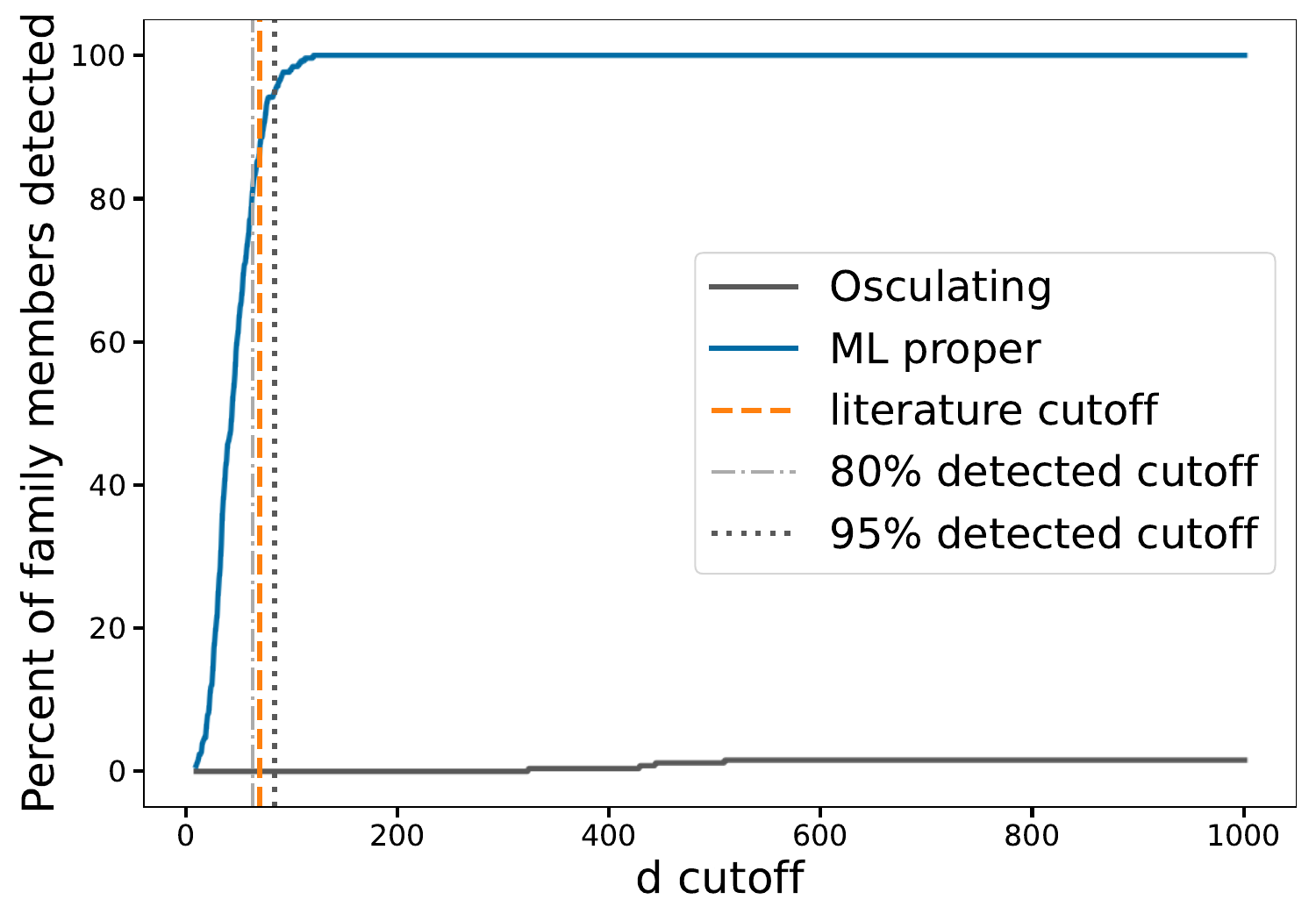}
    \caption{Comparison between using osculating orbital elements and machine learning predicted $\mathrm{e_{prop}}$'s detection rate vs d cutoff chosen relation for the Velleda asteroid family.
    }
    \label{fig:d_detection_rate}
\end{figure}

\begin{figure*}
    \centering
    \includegraphics[width=\textwidth]{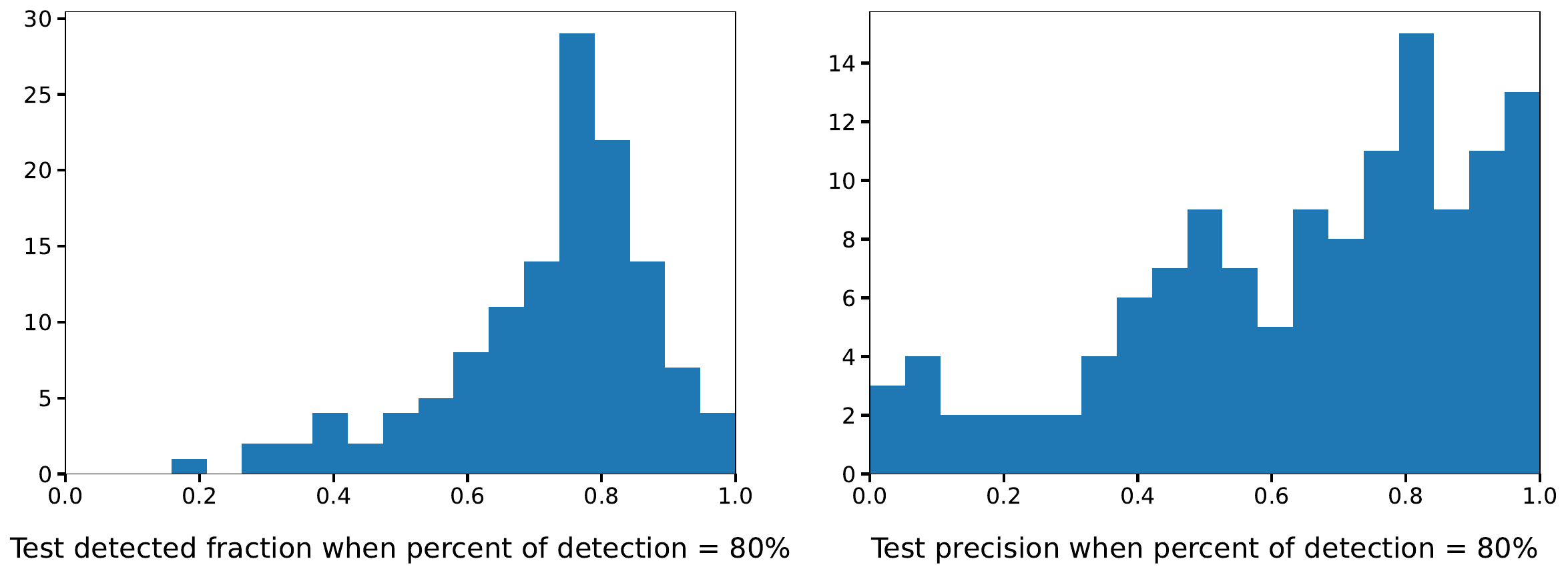}
    \caption{The fraction of detection (recall) in the test set when percent of detection = $80\%$ across 129 asteroids families in the training set, and the precision in the test set when percent of detection = $80\%$ in the training set.}
    \label{fig:precision_ratio_d90}
\end{figure*}
We picked the appropriate $d_\text{cutoff}$ values for each family that ensures $80\%$ and $95\%$ of detection (recall, vertical dot-dash dotted grey lines in Figure\:\ref{fig:Velleda_family}). Because the ML model is not perfect, it is different from the Nesvorn\'y catalog's published $d_\text{cutoff}$ (vertical dashed orange line). Indeed, a better model will in general provide a better precision at a given recall than a worse one (and vice versa).

For example, we compare the precision of our ML model against a toy model that uses osculating orbital elements as proper elements in Figure \ref{fig:d_detection_rate}.
This is the same blue curve as in Figure\:\ref{fig:Velleda_family}, but one must zoom out to much larger $d_\text{cutoff}$ values for the osculating model to encompass a significant number of family members (cf. left panel of Figure\:\ref{fig:family_2d}).
Indeed, at $d_\text{cutoff} = 121$ m/s, the ML model is able to recover 100\% of family members, yet the osculating model recovers only a couple percent.
Only at much larger values of $d_\text{cutoff}$ does the osculating model start recovering a significant fraction of family members, leading to sub-percent precision using osculating values.

Another way to quantify the value of higher precision is by considering a practical strategy where one starts with the full asteroid sample, determines ML-predicted family members, and then only runs computationally expensive N-body proper elements on this reduced subset of ML candidates.
The higher the model's precision, the fewer candidates there are, and the fewer CPU hours must be expended.

In Figure~\ref{fig:d_detection_rate}, we see a much lower $d_\text{cutoff}$ threshold is required to recover $100 \%$ of the family members using the machine learning model compared to osculating elements, which results in many fewer asteroids that need N-body proper elements. Using the candidates identified using machine learning elements, we estimate that $202$ CPU hours ($0.5$ hour per asteroid) are needed to integrate all selected asteroids and confirm their family members. This is approximately $102$ times faster than the estimated $20740$ CPU hours required if one made the selection using osculating elements. We repeated this analysis for 10 families and obtained an average $\sim 90$-fold speedup.

\begin{figure}
    \centering
    \includegraphics[width=0.8\columnwidth]{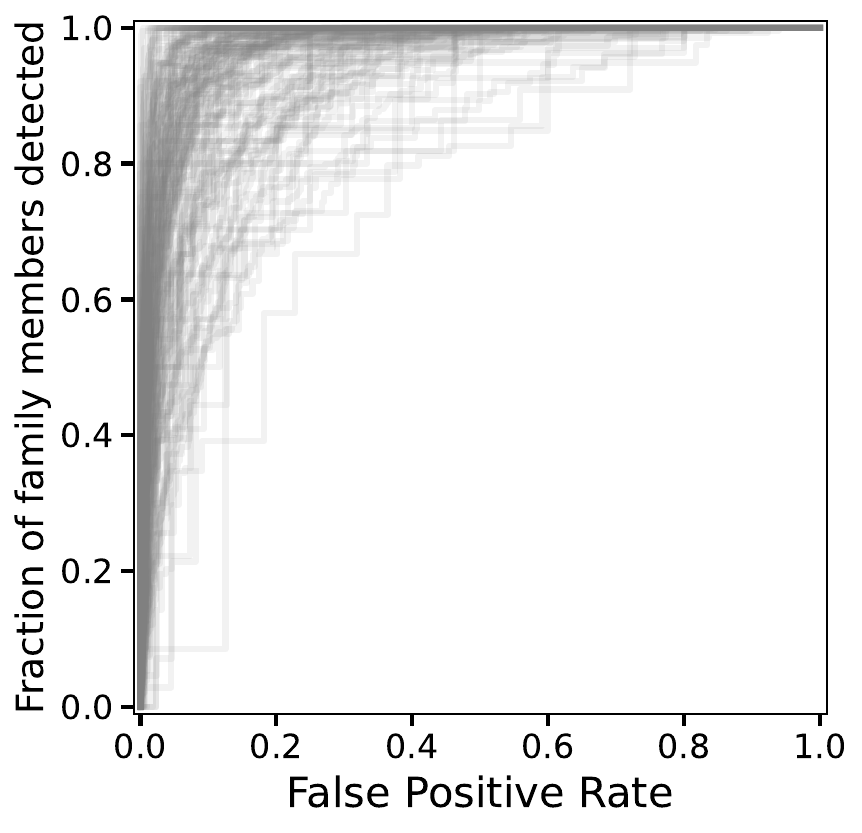}
    \caption{The true/false positive tradeoff (ROC) curve for 153 asteroid families as a function of the selected $d$ threshold. The average area under the curve (AUC) for all 153 families is 0.95 (AUC of 1 indicates a perfect classifier).}
    \label{fig:precision_recall_curve}
\end{figure}

We repeated the analysis from Figure\:\ref{fig:Velleda_family} for all 153 families from \cite{2024ApJS..274...25N}. In Figure\:\ref{fig:precision_recall_curve}, we plot standard ROC curves for our model's performance on each of the 153 families in the train set, i.e., the tradeoff between the false positive rate (how many background asteroids are mislabeled as family members) and the percent of family members detected (recall).
On a given curve, each point corresponds to a different value of $d_\text{cutoff}$ as it is smoothly varied to trade off the $x$ and $y$ quantities.
A perfect model would hug the left and top axes, with a 0\% false-positive rate at 100\% recall. 
A common metric for such classifiers is the Area Under the Curve (AUC), where a perfect model would reach unity.
We find that the average AUC for our ML model across all 153 families is 0.95.

We emphasize that this evaluation is constrained to asteroids within a slab surrounding the known family members. Thus, the ROC curves only characterize how well our model can classify asteroids into nearby background asteroids. In the next section, we measure the model's ability to classify asteroids whose family membership is unknown.

We also determine the $d$ cut-offs corresponding to $80\%$ and $95\%$ percent of detection for each family in the training set. To assess the model's performance on asteroids it had not seen during training, we applied the $80\%$ $d$ cut-offs to test asteroids from the 129 of the 153 families (families with at least $5$ members in the test set for meaningful statistics).\footnote{The following families were excluded: Moore Sitterly, 1994 AO1, Algoa, 2002 FX3, 1999RX33, 1999 CG62, 1999 AE7, Liriope, Kodaihasu, Schlichting, Tuchkova, 2008 YL101, Alamosa, 2002 TE180, 2001 SQ46, Vassar, Bascom, Kazuhiko, 2000 QK75, 2006 QX49, Majella, 1990SB1, Aivazovskij, 1999 VT1.} The right panel of Figure \ref{fig:precision_ratio_d90} shows a histogram of this statistic across the 129 families. The left histogram shows a distribution centered near $0.8$ fraction of detection, meaning that the $d$ cutoffs for $80\%$ of detection rate in the training set usually lead to $\sim 80\%$ in the test set. Additionally, in the right histogram, we show the test sample's precision at $80\%$ $d$ cutoffs. The diagram shows $\sim70\%$ of the families in the test set have precision $> 50\%$ with $80\%$ detection rate in the training set, showing that even though there are tradeoffs, our model still manages to achieve an acceptable precision with high detection rate.

We summarize the $80\%$ and $95\%$ $d$ cutoffs in the training set, the associated precision and percentage of detection in the test set, and the number of test members for each family in an online table (\url{doi.org/10.5281/zenodo.20693707}; \texttt{d\_cutoff\_80\_95\_percent.csv}).

\subsection{Family Identification Pipeline} \label{subsec: Family Identification Pipeline}
As a final test, we evaluate how well our model can assign asteroid family membership on a random asteroid from each of the 129 families with members in the test set, together with an equal number of family-less asteroids. 

For each asteroid, we query the osculating elements, integrate to the JD2460200.5 epoch, then calculate linear and machine learning predicted proper elements.
Since we are obtaining asteroid information through NASA Horizons, the sampled asteroid's designator must start with 'K' and have exactly 7 characters for reliable Horizons querying. 
To limit the search space, we find the five families that the asteroid is closest to and measure the minimum $d$ distance across all members of each family.
We assign the asteroid to the family that it is closest to, unless it not within the 95\% $d$ cutoff of any family, in which case it is considered family-less (see Appendix\:\ref{subsec:appendix pipeline} for a full explanation of the pipeline).

Asteroids known to be in a family are assigned to the correct family 92\% of the time.
$98\%$ of these asteroids that do not belong to any family are classified as such, making the overall accuracy of both test sets $95\%$.

\section{Conclusion}
\label{sec: Conclusion}

In summary, we have trained a \texttt{XGBoost} machine learning model to predict the proper elements of main-belt asteroids with 1.25 million numerically computed objects from \cite{2024ApJS..274...25N}. The model is trained with an input of both the Cartesian elements at JD2460200.5 epoch and a linear estimate of the proper eccentricity and inclination. The \texttt{XGBoost} model with optimal hyperparameters achieved an RMSE of $\eccRMSE$ when comparing to the N-body $e_p$, which is $\approx 2.7 \times$ closer to the proper elements than the osculating values. For $\sin(i_p)$, the model achieved a RMSE value of $\incRMSE$, which is also $\approx 2.2 \times$ closer to the proper elements than the osculating values. The prediction takes $0.74$ CPU hours for 0.25 million asteroids in the test set, $\approx 150,000$ times faster than the approach from \cite{2024ApJS..274...25N} using N-body integration.

We also tested our model on its capability to identify new members of asteroid families. With Velleda family as an example, using model-predicted proper elements as reference, we can detect $87 \%$ of the family members taking the $d$ cutoff from \cite{2024ApJS..274...25N}. 

Across 153 asteroid families, we additionally tabulated the $d$ cutoff required to achieve a $95\%$ detection rate and the precision rate corresponding to this cutoff, in order to guide user trying to balance the tradeoff between percent of detection and precision. Using this cutoff we also showed that 95\% of the time we can identify whether an asteroid is in a family and if so, which family it is in with no prior assumptions about it.

Even if one requires N-body proper elements, our model can reduce the CPU hours needed to confirm asteroids' family membership by factor of $\sim 90$. While not perfect, our model can sufficiently cluster asteroids in proper-element space to reduce the number of candidates to be numerically integrated by two orders of magnitude. 

This work highlights the machine learning model's ability to predict asteroids' proper elements and classify them into families with significantly less CPU hours. Future works might involve developing models that generalize to other regions of phase space (such as TNOs), improving linear theory predictions near resonances, and alternative clustering methods to classify asteroid families.

\section*{Acknowledgments}
\label{sec: Acknowledgments}
We are grateful to Dallin Spencer for helpful comments and discussions.
KV acknowledges funding from NASA grant 80NSSC23K0886. DT acknowledges Cottrell Scholar Award \#CS-CSA-2026-162 from the Research Corporation for Science Advancement. The presented numerical calculations were made possible by computational resources provided through an endowment by the Albrecht family.
 
\appendix



\section{Full pipeline for object family identification}
\label{subsec:appendix pipeline}
In this appendix we present an \href{https://github.com/YangShen31/proper_elements_prediction/blob/main/full_pipeline_example.ipynb}{example notebook} and \href{https://github.com/YangShen31/proper_elements_prediction/blob/main/5_evaluate_full_pipeline.py}{evaluation code} to use our model to predict the proper elements of custom objects and thus further classify them into asteroid family. Firstly, to use our model, which is trained on the epoch Nesvorn\'y's dataset is measured at, one needs the osculating orbital elements at that epoch. 
We provide example code using REBOUND \citep{rebound} to query the NASA Horizons database with an object's designation or manually providing the state vector of the asteroid in REBOUND's default units and integrate it to the JD2460200.5 epoch. We use ASSIST \citep{2023PSJ.....4...69H} on top of REBOUND as an ephemeris integrator for the highest accuracy calculation. Because ASSIST integration works best in the ICRF frame, we first rotate the ecliptic frame state vector into the ICRF frame. Next, we convert the velocities from $\text{AU}/(2\pi \times\text{year})$ to $\text{AU}/\text{day}$ to match the default units of ASSIST. Finally, we integrate from the user's epoch to the Nesvorn\'y epoch using ASSIST's function \texttt{integrate\_or\_interpolate}. We retrieve the state vector in the ICRF frame and convert it back to the J2000 ecliptic frame. Finally, the osculating orbital elements are calculated using REBOUND by specifying the primary object as the Sun and $G=1$. 

We then rotate the resulting osculating elements into the invariable frame and use the rotated elements to estimate the $e_p$ and $i_p$ with linear theory. The linear estimated and Cartesian orbital elements are passed to our model as a ``feature vector'' to predict the proper elements. The pipeline then applies a modified family classification algorithm for each input asteroid. We first use the proper elements of the target asteroid to get five asteroid families that are closest to the asteroid in the proper element space. Then, for each of these five families, we calculate the pairwise distance between family asteroid and the target asteroid and check whether or not the distance is within the Nesvorn\'y published d cutoff. If so, the pipeline identifies the target asteroid to belong to the family.


\section{Linear theory predictions for proper elements}
\label{sec:appendix:linear_theory}
In this appendix we describe our methodology for computing the approximate asteroid proper elements used as input in our ML model via linear secular theory.
Our calculations of proper elements are derived from a Hamiltonian model describing the secular gravitational forcing of the giant planets on a given asteroid.  As a first step in the construction of this model, we develop a synthetic secular theory that describes the giant planets' secular dynamics in Section \ref{sec:appendix:synthetic}.  This quasi-periodic approximation is then used to model the giant planets' forcing of asteroids via a Hamiltonian model, described in Section \ref{sec:appendix:hamiltonian}. 

\subsection{Synthetic Secular Theory for the giant planets}
\label{sec:appendix:synthetic}
The secular evolution of the giant planets' inclinations and eccentricities are, to a good approximation, quasi-periodic \citep{Laskar2008} and we develop a quasi-periodic approximation of their evolution derived from an $N$-body numerical integration.
In our synthetic secular theory, the time dependence of complex dynamical variables
\begin{eqnarray}
X_i&=\sqrt{2(1-\sqrt{1-e_i^2})}\exp\left[\mathrm{i}\varpi_i\right]    
\\
Y_i&=(1-e_i^2)^{1/4}\sin\left(\frac{1}{2}I_i\right)\exp\left[\mathrm{i}\Omega_i\right]~,
\end{eqnarray}
where $e_i$, $I_i$, $\varpi_i$ and $\Omega_i$ are the invariable frame eccentricity, inclination, longitude of perihelion, and longitude of ascending node, respectively, of the $i$th planet are represented as sums of Fourier components,
\begin{align}
    X_i(\boldsymbol{\phi}) &= \sum_{l=1}^{M_i}
    \tilde{X}_{il}\exp\left[\mathrm{i}\mathbf{m}_{l} \cdot {\boldsymbol{\phi}} \right]
    \label{eq:X_fourier}
    \\
    Y_i(\boldsymbol{\phi}) &= \sum_{l=1}^{N_i}
    \tilde{Y}_{il}\exp\left[\mathrm{i}\mathbf{n}_{l} \cdot {\boldsymbol{\phi}} \right],
    \label{eq:Y_fourier}
\end{align}
where $\mathbf{m}_{l}$ and  $\mathbf{n}_{l}$ are vectors of integers, 
$\boldsymbol{\phi} = \boldsymbol{\omega}t + \boldsymbol{\phi}_0$ is a 7-dimensional vector that represents the phases of the systems' secular modes, and $\boldsymbol{\omega} = (g_5,g_6,g_7,g_8,s_6,s_7,s_8)$ is the vector of the giant planet's secular frequencies. 

The frequencies, $\boldsymbol{\omega}$, and Fourier amplitudes $\tilde{X}_{il}$ and $\tilde{Y}_{il}$, are determined by  applying frequency-modified Fourier transform (FMFT) algorithm of \citet{vsidlichovsky1996frequency} to a numerical integration of the outer planets. We use the \texttt{rebound} $N$-body code to integrate the Sun and the four giant planets, added to the simulation with the code's interface for querying the NASA Horizons\footnote{\href{https://ssd.jpl.nasa.gov/horizons/}{ssd.jpl.nasa.gov/horizons/}} database at the JD2460200.5 epoch to align phases of all frequency components, and \texttt{reboundx} for general relativistic effects. Next, we integrate the system for $10\mathrm{~Myr}$, saving the state of the system at equally-spaced intervals of $\sim 1.2\mathrm{kyr}$, generating a total of 8192 save states. We use the \texttt{WHFast} \citep{2015MNRAS.452..376R} implementation of the Wisdom-Holman symplectic integration algorithm \citep{1991AJ....102.1528W} with a time-step of $\sim 173 \mathrm{~d}$, which is $\frac{1}{25}$ of Jupiter's orbital period.

For each planet, we use the FMFT algorithm to determine the frequencies and complex amplitudes of the 14 largest Fourier components of each planet's time series of $X_i$ and $Y_i$. Next we identify the secular mode frequencies $g_i$ and $s_i$ from these numerical Fourier decompositions. The frequencies $g_5,g_6,$ and $g_7$ are set to the frequencies of the three highest-amplitude Fourier components detected in Jupiter's $X_i$ time series and the $g_8$ frequency is set to the largest-amplitude Fourier component of Neptune's $X_i$ time-series. The frequencies $s_6,s_7$ and $s_8$ are set to the largest, third-largest, and second-largest Fourier components detected in Jupiter's $Y_i$ time series.

Fourier components of each planet's $X_i$ and $Y_i$, determined via FMFT, with frequencies matching any of the $g_i$ or $s_i$ frequencies to within $0.1\%$ are assigned to the corresponding secular modes.
After identifying Fourier components associated with these secular modes, we search for additional Fourier components with frequencies of the form $\mathbf{k}\cdot(g_5,\ldots,g_8,s_6,\ldots,s_8)$, satisfying $\sum_i |k_i|=3$ and $\sum_i k_i=1$. The former constraint selects components of the giant planets' quasiperiodic secular evolution that arise as leading nonlinear corrections to linear secular theory. The latter constraint follows from the d'Alembert rules associated with the rotational symmetry of the problem. These rules additionally require that $\sum_{i=5}^{7}k_i\equiv 0\pmod{2}$ for components of $X_i$ and $\sum_{i=5}^{7}k_i\equiv 1\pmod{2}$ for components of $Y_i$.
The selected Fourier components for each planet are shown in Tables \ref{tab:ecc-harmonics} and \ref{tab:inc-harmonics}. In our simulations $(g_5,\ldots,g_8,s_6,\ldots,s_8)=[4.246,28.241,3.089,0.673,-26.338,-2.994,-0.692]$.

\begingroup

\captionsetup[figure]{name={Table},}
\renewcommand{\thefigure}{2}
\begin{figure}[htbp]
\centering

\begin{tabular}{cc}
\begin{subfigure}{0.47\textwidth}
\centering
\scriptsize
\begin{tabular}{lcl}
        Frequency & Amplitude & $\mathbf{k}$ \\ \midrule
+04.246 & 0.0442 $\measuredangle$ 1.82 & \obeyspaces\texttt{[ 1  0  0  0  0  0  0]}\\
+28.241 & 0.0157 $\measuredangle$ 0.52 & \obeyspaces\texttt{[ 0  1  0  0  0  0  0]}\\
+03.089 & 0.00184 $\measuredangle$ -2.91 & \obeyspaces\texttt{[ 0  0  1  0  0  0  0]}\\
+29.398 & 0.000197 $\measuredangle$ -1.04 & \obeyspaces\texttt{[ 1  1 -1  0  0  0  0]}\\
+27.589 & 0.000106 $\measuredangle$ 2.16 & \obeyspaces\texttt{[ 1  0  0  0 -1  1  0]}\\
+52.237 & 0.000571 $\measuredangle$ 2.35 & \obeyspaces\texttt{[-1  2  0  0  0  0  0]}\\
+27.085 & 0.000201 $\measuredangle$ -1.07 & \obeyspaces\texttt{[-1  1  1  0  0  0  0]}
\end{tabular}
\caption{Jupiter}
\end{subfigure}
&
\begin{subfigure}{0.47\textwidth}
\centering
\scriptsize
\begin{tabular}{lll}
        Frequency & Amplitude & $\mathbf{k}$ \\ \midrule
+04.246 & 0.033 $\measuredangle$ 1.82 & \obeyspaces\texttt{[ 1  0  0  0  0  0  0]}\\
+28.241 & 0.0482 $\measuredangle$ -2.63 & \obeyspaces\texttt{[ 0  1  0  0  0  0  0]}\\
+03.089 & 0.00153 $\measuredangle$ -2.91 & \obeyspaces\texttt{[ 0  0  1  0  0  0  0]}\\
+29.398 & 0.000609 $\measuredangle$ 2.11 & \obeyspaces\texttt{[ 1  1 -1  0  0  0  0]}\\
+27.589 & 0.000324 $\measuredangle$ -0.99 & \obeyspaces\texttt{[ 1  0  0  0 -1  1  0]}\\
+52.237 & 0.00193 $\measuredangle$ -0.79 & \obeyspaces\texttt{[-1  2  0  0  0  0  0]}\\
+27.085 & 0.000615 $\measuredangle$ 2.07 & \obeyspaces\texttt{[-1  1  1  0  0  0  0]}
\end{tabular}
\caption{Saturn}
\end{subfigure}
\\[1em]

\begin{subfigure}{0.47\textwidth}
\centering
\scriptsize
\begin{tabular}{lll}
        Frequency & Amplitude & $\mathbf{k}$ \\ \midrule
+04.246 & 0.0381 $\measuredangle$ -1.32 & \obeyspaces\texttt{[ 1  0  0  0  0  0  0]}\\
+28.241 & 0.00155 $\measuredangle$ 0.52 & \obeyspaces\texttt{[ 0  1  0  0  0  0  0]}\\
+03.089 & 0.029 $\measuredangle$ -2.91 & \obeyspaces\texttt{[ 0  0  1  0  0  0  0]}\\
+05.402 & 0.000459 $\measuredangle$ 0.33 & \obeyspaces\texttt{[ 2  0 -1  0  0  0  0]}
\end{tabular}
\caption{Uranus}
\end{subfigure}
&
\begin{subfigure}{0.47\textwidth}
\centering
\scriptsize
\begin{tabular}{lll}
        Frequency & Amplitude & $\mathbf{k}$ \\ \midrule
+04.246 & 0.00194 $\measuredangle$ 1.82 & \obeyspaces\texttt{[ 1  0  0  0  0  0  0]}\\
+28.241 & 0.000103 $\measuredangle$ 0.52 & \obeyspaces\texttt{[ 0  1  0  0  0  0  0]}\\
+03.089 & 0.00371 $\measuredangle$ 0.23 & \obeyspaces\texttt{[ 0  0  1  0  0  0  0]}\\
+00.673 & 0.00912 $\measuredangle$ 2.57 & \obeyspaces\texttt{[ 0  0  0  1  0  0  0]}\\
+05.390 & 5.87e-05 $\measuredangle$ -2.59 & \obeyspaces\texttt{[ 0  0  1  0  0 -1  1]}
\end{tabular}
\caption{Neptune}
\end{subfigure}

\end{tabular}

\caption{\label{tab:ecc-harmonics} Eccentricity harmonics included in our secular theory. Frequency is in units of ''/yr.}
\end{figure}

\renewcommand{\thefigure}{3}
\begin{figure}[htbp]
\centering

\begin{tabular}{cc}
\begin{subfigure}{0.47\textwidth}
\centering
\scriptsize
\begin{tabular}{lcl}
        Frequency & Amplitude & $\mathbf{k}$ \\ \midrule
-26.338 & 0.00315 $\measuredangle$ 0.18 & \obeyspaces\texttt{[ 0  0  0  0  1  0  0]}\\
-02.994 & 0.000485 $\measuredangle$ -2.58 & \obeyspaces\texttt{[ 0  0  0  0  0  1  0]}\\
-00.692 & 0.000574 $\measuredangle$ 1.69 & \obeyspaces\texttt{[ 0  0  0  0  0  0  1]}\\
-26.990 & 2.34e-05 $\measuredangle$ -1.28 & \obeyspaces\texttt{[ 1 -1  0  0  0  1  0]}\\
-01.837 & 1.03e-05 $\measuredangle$ 1.93 & \obeyspaces\texttt{[ 1  0 -1  0  0  1  0]}\\
+82.820 & 1.43e-05 $\measuredangle$ 0.85 & \obeyspaces\texttt{[ 0  2  0  0 -1  0  0]}\\
-02.342 & 1.81e-05 $\measuredangle$ 2.02 & \obeyspaces\texttt{[-1  1  0  0  1  0  0]}\\
-04.150 & 1.02e-05 $\measuredangle$ 2.16 & \obeyspaces\texttt{[-1  0  1  0  0  1  0]}
\end{tabular}
\caption{Jupiter}
\end{subfigure}
&
\begin{subfigure}{0.47\textwidth}
\centering
\scriptsize
\begin{tabular}{lll}
        Frequency & Amplitude & $\mathbf{k}$ \\ \midrule
-26.338 & 0.00785 $\measuredangle$ -2.96 & \obeyspaces\texttt{[ 0  0  0  0  1  0  0]}\\
-02.994 & 0.000394 $\measuredangle$ -2.58 & \obeyspaces\texttt{[ 0  0  0  0  0  1  0]}\\
-00.692 & 0.000553 $\measuredangle$ 1.69 & \obeyspaces\texttt{[ 0  0  0  0  0  0  1]}\\
+34.829 & 1.58e-05 $\measuredangle$ 0.32 & \obeyspaces\texttt{[ 2  0  0  0 -1  0  0]}\\
+58.824 & 1.98e-05 $\measuredangle$ -0.99 & \obeyspaces\texttt{[ 1  1  0  0 -1  0  0]}\\
-26.990 & 5.98e-05 $\measuredangle$ 1.86 & \obeyspaces\texttt{[ 1 -1  0  0  0  1  0]}\\
+82.820 & 3.55e-05 $\measuredangle$ -2.29 & \obeyspaces\texttt{[ 0  2  0  0 -1  0  0]}
\end{tabular}
\caption{Saturn}
\end{subfigure}
\\[1em]

\begin{subfigure}{0.47\textwidth}
\centering
\scriptsize
\begin{tabular}{lll}
        Frequency & Amplitude & $\mathbf{k}$ \\ \midrule
-26.338 & 0.000354 $\measuredangle$ 0.18 & \obeyspaces\texttt{[ 0  0  0  0  1  0  0]}\\
-02.994 & 0.0089 $\measuredangle$ 0.56 & \obeyspaces\texttt{[ 0  0  0  0  0  1  0]}\\
-00.692 & 0.000555 $\measuredangle$ -1.46 & \obeyspaces\texttt{[ 0  0  0  0  0  0  1]}\\
-01.837 & 0.000182 $\measuredangle$ -1.08 & \obeyspaces\texttt{[ 1  0 -1  0  0  1  0]}\\
-02.342 & 0.000298 $\measuredangle$ -1.12 & \obeyspaces\texttt{[-1  1  0  0  1  0  0]}\\
-01.185 & 2.62e-05 $\measuredangle$ 0.47 & \obeyspaces\texttt{[ 0  1 -1  0  1  0  0]}\\
-04.150 & 0.000194 $\measuredangle$ -0.96 & \obeyspaces\texttt{[-1  0  1  0  0  1  0]}
\end{tabular}
\caption{Uranus}
\end{subfigure}
&
\begin{subfigure}{0.47\textwidth}
\centering
\scriptsize
\begin{tabular}{lll}
        Frequency & Amplitude & $\mathbf{k}$ \\ \midrule
-26.338 & 3.78e-05 $\measuredangle$ 0.18 & \obeyspaces\texttt{[ 0  0  0  0  1  0  0]}\\
-02.994 & 0.00107 $\measuredangle$ -2.57 & \obeyspaces\texttt{[ 0  0  0  0  0  1  0]}\\
-00.692 & 0.00578 $\measuredangle$ -1.46 & \obeyspaces\texttt{[ 0  0  0  0  0  0  1]}\\
-02.342 & 5.6e-05 $\measuredangle$ 2.02 & \obeyspaces\texttt{[-1  1  0  0  1  0  0]}\\
-04.150 & 2.89e-05 $\measuredangle$ 2.22 & \obeyspaces\texttt{[-1  0  1  0  0  1  0]}
\end{tabular}
\caption{Neptune}
\end{subfigure}

\end{tabular}

\caption{\label{tab:inc-harmonics} Inclination harmonics included in our secular theory. Frequency is in units of ''/yr.}
\end{figure}
\endgroup 

\subsection{Asteroid Secular Hamiltonian}
\label{sec:appendix:hamiltonian}
We formulate the Hamiltonian describing the dynamics of a given asteroid in terms of the complex canonical variables 
\begin{align}
    x&=\sqrt{2}\sqrt{1-\sqrt{1-e^2}}\exp[\mathrm{i}\varpi]
    \\
    y&=2(1-e^2)^{1/4}\sin(I/2)\exp[\mathrm{i}\Omega],
\end{align}
their complex conjugates, $\bar{x}$ and $\bar{y}$, the secular mode phase variables $\boldsymbol{\phi}=(\phi_1,....,\phi_7)$, and conjugate ``dummy" actions $\boldsymbol{\Phi}=(\Phi_1,....,\Phi_7)$. In terms of these variables, the Hamiltonian, which captures the secular interactions between the asteroid and planets to second order in eccentricities and inclinations, reads
\begin{equation}
H(x,y,\bar{x},\bar{y},{\boldsymbol{\Phi}},\boldsymbol{\phi}) =  -g_0|x|^2 - s_0|y|^2 
   +\boldsymbol{\omega}\cdot  \boldsymbol\Phi
    + 
    x\bar{\mathcal{X}}(\boldsymbol{\phi}) 
    +
    \bar{x}{\mathcal{X}}(\boldsymbol{\phi})
    +
    y\bar{\mathcal{Y}}(\boldsymbol{\phi}) 
    +
    \bar{y}{\mathcal{Y}}(\boldsymbol{\phi})
    \label{eq:hamiltonian}
\end{equation}
where $g_0$ and $s_0$ are the secular free precession frequencies, $\mathcal{X}(\phi)$ and $\mathcal{Y}(\phi)$ represent the forcing terms of the planets, and overbars denote complex conjugates.
The equations of motion for the Hamiltonian  (Equation \ref{eq:hamiltonian}) are given by
\begin{align}
\dd{x}{t} &= -\mathrm{i}\partial_{\bar{x}}H \label{eq:xdot}
\\
\dd{y}{t} &= -\mathrm{i}\partial_{\bar{y}}H \label{eq:ydot}
\\
\dd{\phi_i}{t}&= \omega_i \label{eq:phidot}~. 
\end{align}

For an asteroid with semi-major axis $a$ and corresponding mean motion, $n$, the free precession frequencies, with disturbing function coefficients written in \texttt{celmech} notation \citep{hadden_celmech_2022}, are 
\begin{align}
    g_0 &= 2n \left(\sum_{i\in\{l| a_l>a\}}\fracbrac{m_i a}{M_\odot a_i}C_\mathbf{0}^{(0,0,1,0)}(a/a_i)
    + \sum_{i\in\{l| a_l<a\}}\fracbrac{m_i}{M_\odot}C_\mathbf{0}^{(0,0,0,1)}(a_i/a)\right)
\\
    s_0 &= \frac{1}{2}n \left(\sum_{i\in\{l| a_l>a\}}\fracbrac{m_i a}{M_\odot a_i}C_\mathbf{0}^{(1,0,0,0)}(a/a_i)
    + \sum_{i\in\{l| a_l<a\}}\fracbrac{m_i}{M_\odot}C_\mathbf{0}^{(0,1,0,0)}(a_i/a)\right)~.
\end{align}
and the coupling terms are
\begin{align}
\label{eq:Xforcing}
  \mathcal{X}(\phi)
  = 
  -n \left(\sum_{i\in\{l| a_l>a\}}\fracbrac{m_i a}{M_\odot a_i}C_{(0,0,1,-1,0,0)}^{\mathbf{0}}(a/a_i)
  X_i(\phi)
    + \sum_{i\in\{l| a_l<a\}}\fracbrac{m_i}{M_\odot}C_{(0,0,1,-1,0,0)}^{\mathbf{0}}(a_i/a) X_i(\phi)\right)
\\
\label{eq:Yforcing}
\mathcal{Y}(\phi)
  = 
  -\frac{n}{2} \left(\sum_{i\in\{l| a_l>a\}}\fracbrac{m_i a}{M_\odot a_i}C_{(0,0,0,0,1,-1)}^{\mathbf{0}}(a/a_i)
  Y_i(\phi)
    + \sum_{i\in\{l| a_l<a\}}\fracbrac{m_i}{M_\odot}C_{(0,0,0,0,1,-1)}^{\mathbf{0}}(a_i/a) Y_i(\phi)\right)~.
\end{align}

These equations can be solved analytically. In order to do so, we first introduce 
Fourier coefficients, $\tilde{\mathcal{X}}_l$ and $\tilde{\mathcal{Y}}_l$, of the forcing terms defined so that 
\begin{align}
    \mathcal{X}(\phi) = \sum_l \tilde{\mathcal{X}}_l \exp[\ii \mathbf{m}_l \cdot \boldsymbol{\phi}]
    \\
    \mathcal{Y}(\phi) = \sum_l \tilde{\mathcal{Y}}_l \exp[\ii \mathbf{n}_l \cdot \boldsymbol{\phi}]~.
\end{align}
It is straightforward to work out explicit expressions for the $\tilde{\mathcal{X}}_l$ and $\tilde{\mathcal{Y}}_l$ by substituting into Equations \eqref{eq:Xforcing} and \eqref{eq:Yforcing} the Fourier decompositions of the planets' $X_i$ and $Y_i$ constructed in  Section \ref{sec:appendix:synthetic}.
Inserting these expressions into Equations \eqref{eq:xdot} and \eqref{eq:ydot}, we find that 
\begin{eqnarray}
    \dd{x}{t}  &= \ii g_0x - \ii \sum_l \tilde{\mathcal{X}}_l \exp[\ii \mathbf{m}_l \cdot \boldsymbol{\phi}]
    \label{eq:xdot_explicit}
    \\
    \dd{y}{t}  &= \ii s_0y - \ii \sum_l \tilde{\mathcal{Y}}_l \exp[\ii \mathbf{n}_l \cdot \boldsymbol{\phi}]
    \label{eq:ydot_explicit}
\end{eqnarray}
Inserting the solution $\boldsymbol{\phi}(t) = \boldsymbol{\omega} t + \boldsymbol{\phi}_0$ of Equation \eqref{eq:phidot} into Equations \eqref{eq:xdot_explicit} and \eqref{eq:ydot_explicit}, we obtain the general solutions
\begin{align}
    x(t) = x_\mathrm{proper}\exp[\ii g_0 t] +  \sum_l \frac{\tilde{\mathcal{X}}_l}{g_0 -\mathbf{m}_l \cdot \boldsymbol{\omega}} \exp[\ii \mathbf{m}_l \cdot \boldsymbol{\phi}(t)]
    \label{eq:x_of_t_linear}
    \\
    y(t) = y_\mathrm{proper}\exp[\ii s_0 t] + \sum_l \frac{\tilde{\mathcal{Y}}_l}{s_0 -\mathbf{n}_l \cdot \boldsymbol{\omega}} \exp[\ii \mathbf{n}_l \cdot \boldsymbol{\phi}(t)]~.
    \label{eq:y_of_t_linear}
\end{align}

Proper eccentricities and inclinations are computed fom $|x_{\text{proper}}|, |y_\text{proper}|$ according to 
\begin{eqnarray}
    e_p &=& \sqrt{\text{clip}\left(\lvert x_\mathrm{proper}\rvert^2 - \frac{1}{4}(\lvert x_\mathrm{proper}\rvert^4\right)}    \\
    \sin(i_p)&=& 2\sin^{-1} \left(\text{clip}\left(\frac{|y_{\text{proper}}|}{2(1-e^2)^{1/4}}\right) \right)
\end{eqnarray}
after solving Equations \eqref{eq:x_of_t_linear} and \eqref{eq:y_of_t_linear} for $x_\mathrm{proper}$ and 
$y_\mathrm{proper}$ at $t=0$. $\text{clip}()$ indicates clipping a value to the range $[0,1]$.
The denominators appearing in our Equations \eqref{eq:x_of_t_linear} and \eqref{eq:y_of_t_linear} diverge at secular resonances and the linear theory approximations become invalid in the vicinity of these resonances. 
We do attempt to correct our estimates of (near-)resonant asteroids' proper secular elements and simply include them in asteroids' feature vectors. While we do not have insight into how the model is using this data, we note that very large linear predictions carry the information that the asteroid is near a secular resonance that in principle can be exploited by the model. Indeed, we can see in Fig.\:\ref{fig:dsini_de_plots} that the model is able to successfully predict the mean elements of near-resonant asteroids that have very large linear predicted values.

\bibliography{main.bib}
\bibliographystyle{aasjournal}

\end{CJK*}
\end{document}